\documentclass[a4paper,11pt]{article}

\usepackage{jcappub} 

\usepackage[T1]{fontenc} 
\usepackage{booktabs}
\usepackage{subcaption}
\usepackage{amsmath}

\title{\boldmath 
Inflationary Magnetogenesis with $f(R,\phi)$ Coupling}

\author[a]{Shuang Liu}
\author[a]{Bo-yu Zhao}
\author[a,1]{Yu Li,\note{Corresponding author.}}
\author[a]{Yao-chuan Wang}

\affiliation[a]{School of Science, Dalian Maritime University, \\Dalian 116026, China}

\emailAdd{1120240893shuangliu@dlmu.edu.cn}
\emailAdd{bo-yuzhao@dlmu.edu.cn}
\emailAdd{leeyu@dlmu.edu.cn}
\emailAdd{ycwang@dlmu.edu.cn}

\abstract{Inflationary magnetogenesis provides a promising mechanism for generating primordial large-scale magnetic fields, but faces challenges such as the strong coupling problem and backreaction issues. In this paper, we extend the Ratra model by introducing a coupling between the electromagnetic field and the background geometry, parameterized as $K(R)I^2(\phi)$. Starting from a general action with $f^2(R,\phi)F_{\mu\nu}F^{\mu\nu}$, we adopt $f^2(R,\phi)=K(R)I^2(\phi)$ as a concrete realization, where $K(R)=1-(R/M^2)^n$ is a curvature-dependent coupling function that deviates from unity only when $R$ becomes comparable to $M^2$. Rather than focusing on the slow-roll inflationary stage (which reduces to the standard Ratra scenario), we concentrate on the post-inflationary reheating epoch, where the broken-power-law evolution of the scale factor and coupling function across the inflation-to-reheating transition allows us to derive analytic expressions for the magnetic and electric energy density spectra. Three key theoretical constraints are imposed on the model parameter space: the strong coupling condition, the backreaction constraint, and the CMB isotropy requirement. Through numerical calculations, we obtain predictions for the present-day magnetic field strength $B_0$ and coherence length $L_{c0}$ for various combinations of the inflationary energy scale $H_f$ and the reheating temperature $T_r$. A key feature of this work is
the incorporation of nonlinear magnetohydrodynamic (MHD) turbulence during the reheating era, including the inverse transfer process, which significantly enhances the coherence length (up to $\sim 0.1$ Mpc) while modifying the field strength by several orders of magnitude. By comparing with observational constraints from radio observations and Fermi-LAT gamma-ray data, we demonstrate that the inflationary energy scale $H_f$, the reheating temperature $T_r$, the parameter $\beta$, and the e-folding numbers $N_f$, $N_r$ must satisfy stringent
joint constraints. This work provides a viable theoretical framework for inflationary magnetogenesis that simultaneously satisfies theoretical consistency conditions and current observational bounds, with the reheating-stage nonlinear MHD evolution serving as a crucial ingredient for producing observationally compatible magnetic fields.
}
\keywords{inflationary magnetogenesis, magnetohydrodynamics (MHD), reheating, primordial magnetic fields}

\begin{document}
\maketitle
\flushbottom
\section{Introduction}
Observational evidence suggests the existence of weak large-scale magnetic fields in the universe, with typical strengths in the range $10^{-15}\,\mathrm{G} \lesssim B_0 \lesssim 10^{-9}\,\mathrm{G}$ \cite{1994_Kronberg,1999_blasi,2010_neronov,2020_osullivan,2020_tiede}. Despite years of extensive research, the origin of these cosmic magnetic fields remains an open and widely debated question. Currently, proposed magnetogenesis mechanisms can be broadly classified into two categories: astrophysical and primordial origins. Astrophysical processes, such as supernova explosions and galactic dynamo effects, can amplify seed magnetic fields on galactic scales, but they struggle to account for the magnetic fields observed in cosmic voids \cite{2002_widrow,2005_hanayama,2018_safarzadeh,2021_araya,2023_papanikolaou,2023_papanikolaoua}. Therefore, magnetic fields in voids are more likely to have a primordial origin, dating back to physical processes in the very early universe \cite{2010_neronov,2017_archambault,2018_ackermann}.

Among various primordial mechanisms, inflationary magnetogenesis stands out as a compelling mechanism \cite{1988_turner,1992_ratra,2008_martin,2014_kobayashi,2015_giovannini,2015_tasinato,2016_domenech,2017_sharma,2018_sharma,2019_shtanov,2021_bamba,2021_giovannini,2022_durrer,2023_velasquez,2023_li,2025_dimopoulos}. During inflation, quantum vacuum fluctuations of the electromagnetic field can be stretched to superhorizon scales, potentially seeding large-scale magnetic fields. However, the standard electromagnetic action is conformally invariant, and the Friedmann-Robertson-Walker (FRW) metric is conformally flat, which prevents the amplification of electromagnetic fields during inflation. Therefore, breaking conformal invariance is a necessary condition for any viable inflationary magnetogenesis model (see \cite{2010_subramanian,2011_kandus,2016_subramanian} for comprehensive reviews).

A widely studied approach to breaking conformal invariance is the Ratra model \cite{1992_ratra}, which introduces a coupling of the form $f^2(\phi)F_{\mu\nu}F^{\mu\nu}$ between the inflaton field $\phi$ and the electromagnetic field tensor $F_{\mu\nu}$, where $f(\phi)$ is a time-dependent coupling function. The Ratra model provides a simple and elegant framework, but it faces several well-known challenges. First, when the coupling function becomes too small during inflation, the strong coupling problem arises, rendering the effective electromagnetic coupling excessively large \cite{2014_kobayashi,2017_sharma}. Second, when the generated electromagnetic energy density exceeds the background inflationary energy density, the backreaction problem occurs, which may spoil the inflationary dynamics \cite{2009_vittoria,2013_giovannini}. Third, the produced magnetic field must survive the subsequent cosmic evolution without being diluted below the observable level.

Various attempts have been made to overcome these difficulties. Earlier works explored electromagnetic-curvature couplings beyond the standard Ratra setup \cite{2008_bamba}. More recently, studies on $f(R)$ curvature-dependent matter coupling during inflation \cite{2026_barroso} have shown that geometric-matter interaction can stabilize inflationary dynamics and avoid the strong-coupling difficulty for negative-power-law $f(R)$ models. Inspired by these developments, we consider a general action with $f^2(R,\phi)F_{\mu\nu}F^{\mu\nu}$ in the electromagnetic sector, and adopt $f^2(R,\phi)=K(R)I^2(\phi)$ as a concrete realization, where $K(R)=1-(R/M^2)^n$ is a curvature-dependent coupling function that deviates from unity only when $R$ becomes comparable to $M^2$. This framework extends the Ratra scenario by introducing curvature-dependent coupling while maintaining a separable structure between curvature and scalar-field sectors.

In this paper, we investigate inflationary magnetogenesis within a $f(R,\phi)$ coupling framework, with particular emphasis on the reheating epoch, where nonlinear magnetohydrodynamic (MHD) evolution plays a crucial role. In Sec.~\ref{s2}, we establish a general coupling framework, present the Maxwell equations within the Lagrangian formalism, and derive analytic expressions for the magnetic and electric energy density spectra. The de Sitter inflationary stage reduces to the standard Ratra scenario, while the reheating transition provides the key novel predictions. In Sec.~\ref{s3}, we systematically study the reheating dynamics and impose three theoretical constraints: the strong coupling condition, the backreaction constraint, and the CMB isotropy requirement. Through detailed numerical calculations, we obtain predictions for the present-day magnetic field strength $B_0$ and coherence length $L_{c0}$ for various combinations of the inflationary energy scale $H_f$ and the reheating temperature $T_r$. A key highlight is the incorporation of nonlinear MHD turbulence during the reheating era, including the inverse transfer process, which significantly modifies the field strength and coherence length. We compare our results with observational constraints from radio observations and Fermi-LAT gamma-ray data. Finally, our conclusions are presented in Sec.~\ref{s4}.

\section{The $f(R,\phi)$ coupling framework \label{s2}}
\subsection{Model setup and equations of motion}
We consider a general coupling framework in which the electromagnetic field interacts with both curvature and a scalar field through a function $f^2(R,\phi)$. Throughout this work we adopt the Jordan frame (for discussions on frame equivalence in inflationary magnetogenesis, see ~\cite{2026_wang}). The action is
\begin{equation}
\label{e1}
S = \int d^4x \sqrt{-g} \left[ \frac{R}{2} - \frac{1}{16\pi} f^2(R,\phi) F_{\mu\nu}F^{\mu\nu} - \frac{1}{2}g^{\mu\nu}\partial_{\mu}\phi\partial_{\nu}\phi - V(\phi) \right]
\end{equation}
For the coupling function we adopt a separable realization
\begin{equation}
\label{e2}
f^2(R,\phi) = K(R)\,I^2(\phi), \qquad 
K(R) = 1 - \left(\frac{R}{M^2}\right)^n
\end{equation}
with $M$ a mass-scale parameter and $n$ a positive integer. Here $g$ is the determinant of the metric, $F_{\mu\nu} \equiv \partial_{\mu}A_{\nu} - \partial_{\nu}A_{\mu}$ is the electromagnetic field tensor, and $\phi$ is the inflaton field with potential $V(\phi)$.

We adopt a spatially flat Friedmann--Robertson--Walker (FRW) metric,
\begin{equation}
\label{e3}
\begin{aligned}
ds^2 &= -dt^2 + a^2(t) \big[ dx^2 + dy^2 + dz^2 \big] \\
     &= a^2(\eta) \big[ -d\eta^2 + dx^2 + dy^2 + dz^2 \big]
\end{aligned}
\end{equation}
where $\eta = \int dt/a(t)$ is the conformal time. In a perturbed FRW background, the temporal component of Maxwell's equations imposes an additional consistency condition on the coupling function~\cite{2022_li,2025_li}; however, for the homogeneous and isotropic background considered here, this constraint is automatically satisfied and does not restrict the coupling.Varying the action \eqref{e1} with respect to $A_\nu$ yields the modified Maxwell equation
\begin{equation}
\label{e4}
\big[ f^2(R,\phi) F^{\mu\nu} \big]_{;\nu} = 0
\end{equation}
In the Coulomb gauge ($A_0 = 0$, $\partial_i A_i = 0$), we expand the vector potential in Fourier modes and define $\bar{A}(\eta,k) \equiv a(\eta) A(\eta,k)$. We define $f(R,\phi) > 0$ as the positive root, ensuring the total coupling
amplitude remains positive throughout the evolution. The equation of motion for $\bar{A}$ then reads
\begin{equation}
\label{e5}
\bar{A}'' + 2\frac{f'}{f} \bar{A}' + k^2 \bar{A} = 0
\end{equation}
where a prime denotes $d/d\eta$. Introducing the canonically normalized field $\mathcal{A} = f(R,\phi)\,\bar{A}$, Eq.~\eqref{e5} is recast into a Schrödinger-like form without the first-derivative term:
\begin{equation}
\label{e6}
\mathcal{A}'' + \left[ k^2 - \frac{f''}{f} \right] \mathcal{A} = 0
\end{equation}
This is the master equation governing the evolution of electromagnetic quantum fluctuations in our $f(R,\phi)$ coupling framework.

\subsection{Energy-momentum tensor and spectral densities}

In our curvature-dependent $f(R,\phi)$ coupling framework, the electromagnetic Lagrangian is $\mathcal{L}_{\rm EM}= - (1/16\pi)f^2(R,\phi)F_{\mu\nu}F^{\mu\nu}$. Varying the total action with respect to the metric yields the electromagnetic energy-momentum 
tensor (EMT). Since $f^2(R,\phi)$ depends on the Ricci scalar $R$, the metric variation receives 
two contributions:
\begin{equation}
\label{e7}
T_{\mu\nu}=T_{\mu\nu}^{(\rm F)}+T_{\mu\nu}^{(\rm Geo)}
\end{equation}
The first term originates from the explicit $F_{\mu\nu}F^{\mu\nu}$ dependence:
\begin{equation}
\label{e8}
T_{\mu\nu}^{(\rm F)} = \frac{f^2(R,\phi)}{4\pi}
\left[ g^{\gamma\beta} F_{\mu\gamma} F_{\nu\beta} - g_{\mu\nu} \frac{F_{\alpha\beta}F^{\alpha\beta}}{4} \right]
\end{equation}
The second term is the geometric contribution arising from the dependence of $f^2(R,\phi)$ on $R$:
\begin{equation}
\label{e9}
T_{\mu\nu}^{(\rm Geo)}= \frac{1}{8\pi} \bigl(R_{\mu\nu}+g_{\mu\nu}\Box-\nabla_\mu\nabla_\nu\bigr)
\bigl[I^2(\phi)K_R\,F_{\alpha\beta}F^{\alpha\beta}\bigr]
\end{equation}
where $K_R=\partial K(R)/\partial R$ and $\Box=g^{\alpha\beta}\nabla_\alpha\nabla_\beta$.

To compute the magnetic and electric energy densities, we start from the $00$-component of 
$T_{\mu\nu}^{(\rm F)}$.  Following the discussion in \cite{2008_bamba}, on a fixed length scale 
$L=2\pi/k$ (where $k$ is the wavenumber), the squared norms of the electric and magnetic fields 
are expressed as
\begin{equation}
\label{e10}
|E_i(\eta)|^2 = \frac{|E_0|^2}{(f^2)^2 a^4}, \qquad |B_i(\eta)|^2 = \frac{|B_0|^2}{a^4}
\end{equation}
where $|E_0|$ and $|B_0|$ are constants. Consequently,
\begin{align}
F_{\alpha\beta}F^{\alpha\beta} &= 2\bigl(|B_i|^2-|E_i|^2\bigr)\label{e11}\\
(F_{\alpha\beta}F^{\alpha\beta})' &= 8\Biggl[\left(\mathcal{H}+\frac{f'}{f}\right)|E_i|^2
-\mathcal{H}|B_i|^2\Biggr]\label{e12}
\end{align}
with $\mathcal{H}=a'/a$.

Substituting these into $T_{00}/a^2$, separating magnetic and electric parts, performing a Fourier 
expansion and taking the vacuum expectation value (VEV) under Bunch--Davies initial conditions, the spectral energy densities 
are obtained from the mode function $\mathcal{A}(k,\eta)$ as \cite{2008_martin,2010_subramanian}
\begin{align}
\frac{{\rm d}\rho_B(k,\eta)}{{\rm d}\ln k} &= \frac{k^5}{2\pi^2 a^4}\frac{F(\eta)}{f^2}|\mathcal{A}|^2\label{e13}\\
\frac{{\rm d}\rho_E(k,\eta)}{{\rm d}\ln k} &= \frac{k^3}{2\pi^2 a^4}G(\eta)\left|\left(\frac{\mathcal{A}}{f}\right)'\right|^2\label{e14}
\end{align}
The functions $F(\eta)$ and $G(\eta)$ are given by
\begin{align}
F(\eta) &= f^2 + \frac{6}{a^2}\Big[ \mathcal{H}\,(I^2 K_R)' - \mathcal{H}' I^2 K_R - 4\mathcal{H}^2 I^2 K_R \Big]\label{e15}\\
G(\eta) &= f^2 - \frac{6}{a^2}\Big[ \mathcal{H}(I^2 K_R)' - \mathcal{H}' I^2 K_R - 4\mathcal{H}^2 I^2 K_R - 4\mathcal{H} I^2 K_R \frac{f'}{f} \Big]\label{e16}
\end{align}
When evaluating the VEV of the energy density associated with the magnetic part of the electromagnetic field, the $R$-dependence of $K(R)$ also generates an additional geometric term $-\frac{1}{8\pi a^2}\nabla^2(I^2 K_R F_{\alpha\beta}F^{\alpha\beta})$, which is independent of the electromagnetic mode functions. Moreover, this term vanishes upon taking the VEV, and therefore does not affect the final magnetic or electric power spectra. Hence, it will not be discussed further.

\subsection{Near-dS approximation, mode functions, and magnetogenesis constraints}

We assume the total coupling scales as
\begin{equation}
f(R,\phi)\propto a^{\alpha}\propto (-\eta)^{\gamma},
\qquad \gamma=\alpha(-1+\delta)\label{e17}
\end{equation}
The scale factor evolves as $a(\eta) \propto (-\eta)^{1+\delta}$, and the Ricci scalar in the Jordan frame is $R=12H^2+6\dot{H}$. The case $\delta=-2$ corresponds to exact de Sitter; we consider the near-de Sitter regime $\delta=-2+\epsilon$ with $\epsilon\ll1$ the slow-roll parameter, yielding $\gamma\approx\alpha(-1+\epsilon)$.

Since $f(R,\phi)\propto(-\eta)^\gamma$, we have $f''/f=\gamma(\gamma-1)/\eta^2$, and the mode function $\mathcal{A}(k,\eta)$ satisfies
\begin{equation}
\label{e18}
\mathcal{A}'' + \left(k^2 - \frac{\gamma(\gamma-1)}{\eta^2}\right)\mathcal{A} = 0
\end{equation}
Deep inside the Hubble radius ($-k\eta\to\infty$), we impose the Minkowski vacuum initial condition $\mathcal{A}\to e^{-ik\eta}/\sqrt{2k}$; in the super-Hubble limit ($-k\eta\to 0$), the solution behaves as
\begin{equation}
\label{e19}
\mathcal{A} \simeq C_1(-k\eta)^{\gamma} + C_2(-k\eta)^{1-\gamma}
\end{equation}
with $C_{1,2}$ fixed by matching to the Bessel-function solution. Substituting \eqref{e19} into \eqref{e13} and \eqref{e14} yields the late-time power spectra:
\begin{align}
\mathcal{P}_B(k) &= \frac{H^4}{2\pi^2}\frac{F(\eta)}{f^2}\,\mathcal{F}(m)\,(-k\eta)^{2m+4}\label{e20}\\
\mathcal{P}_E(k) &= \frac{H^4}{2\pi^2}\frac{G(\eta)}{f^2}\,\mathcal{G}(l)\,(-k\eta)^{2l+4}\label{e21}
\end{align}
where
\begin{align}
\mathcal{F}(m) &= \frac{\pi}{2^{2m+1}\cos^2(m\pi)\,\Gamma^2(m+\frac12)}\label{e22}\\
\mathcal{G}(l) &= \frac{\pi}{2^{2l+1}\cos^2(l\pi)\,\Gamma^2(l+\frac12)}\label{e23}
\end{align}
with $m=\gamma$ for $\gamma\le1/2$ ($m=1-\gamma$ otherwise) and $l=\gamma+1$ for $\gamma\le-1/2$ ($l=-\gamma$ otherwise).

Under the near-de Sitter approximation, both ratios $F(\eta)/f^2$ and $G(\eta)/f^2$ approach unity and will be set to 1 in the following analysis.

For a scale-invariant magnetic spectrum, the coupling $f(R,\phi)\propto a^\alpha$ admits two mathematical solutions: $\alpha=-3$ and $\alpha=2$.  The $\alpha=-3$ branch leads to a rapid growth of the electric energy density, causing a severe backreaction problem that disrupts the inflationary background, and is therefore physically disallowed.  The $\alpha=2$ branch avoids backreaction \cite{2017_sharma} but suffers from the strong coupling problem: at early times $f$ becomes too small, making the effective electromagnetic coupling $e_{\rm eff}=e/f$ too large for perturbation theory to be valid. 

During de Sitter inflation, the Ricci scalar $R = 12H^2$ is constant.
Whether $K(R) = 1 - (R/M^{2})^{n}$ deviates appreciably from unity depends on
the value of $M$ relative to $H_{\text{inf}}$. For $M \gtrsim M_{\text{Pl}}$, $R/M^{2} \ll 1$ during inflation and $K(R) \approx 1$ at leading order,
recovering the classic Ratra scenario and explaining why the near-dS analysis above yields the standard $\alpha = 2$ conclusion. The curvature-dependent modification becomes relevant only at the reheating stage where $R$ grows significantly and the coupling deviates from unity. This reheating-era
modification is the central innovation of this work, as it circumvents the strong coupling problem while preserving the advantages of the $\alpha = 2$
branch. We develop this modified model in detail in section~\ref{s3}.

\section{Dynamics and Observational Constraints during Reheating \label{s3}}

The electromagnetic sector is described by
\begin{equation}
\label{e31}
S_{Em} = -\frac{1}{16\pi} \int d^4x \sqrt{-g} f^2(R,\phi) F_{\mu\nu}F^{\mu\nu}
\end{equation}
where the coupling $f^2(R,\phi)$ follows the separable form in Eq.~\eqref{e2} and depends on conformal time through $R(\eta)$ and $\phi(\eta)$.

The background evolution is described by a broken-power-law scale factor across the inflation-to-reheating transition:
\begin{equation}
\label{e32}
a(\eta) = 
\begin{cases} 
-\dfrac{1}{H_{\text{f}}\eta} & \eta_i \le \eta \le \eta_{\text{f}} \\
\dfrac{a_{\text{f}}^3 H_{\text{f}}^2}{4} \left( \dfrac{3}{a_{\text{f}}H_{\text{f}}} + \eta \right)^2 & \eta_{\text{f}} \le \eta \le \eta_{\text{r}}
\end{cases}
\end{equation}
During inflation ($\eta_i \le \eta \le \eta_{\text{f}}$), the universe is approximately de Sitter; during reheating ($\eta_{\text{f}} \le \eta \le \eta_{\text{r}}$), the scale factor follows a quadratic evolution $a\propto\eta^2$, corresponding to a matter-dominated-like expansion. The field factor $I^2(\phi)$ is defined piecewise as
\begin{equation}
\label{e33}
I^2(\phi) = 
\begin{cases} 
\left(\dfrac{a}{a_i}\right)^{\nu_1} & a_i \le a \le a_{\text{f}} \\
\left(\dfrac{a_{\text{f}}}{a_i}\right)^{\nu_1} \left(\dfrac{a}{a_{\text{f}}}\right)^{\nu_2} & a_{\text{f}} \le a \le a_{\text{r}}
\end{cases}
\end{equation}
As discussed in Sec.~\ref{s2}, we adopt the $\alpha=2$ branch, since the electric energy density does not diverge during inflation, thereby avoiding the severe backreaction problem of the $\alpha=-3$ branch. The trade-off is the strong-coupling problem at very early times.

We assume the initial coupling constant at the onset of inflation to be $f_i=1$. Under this assumption, $f$ attains a large value by the end of inflation, rendering the effective coupling constant extremely small, which is acceptable in the early universe~\cite{2010_subramanian}. To address the possible backreaction problem caused by the rapid decay of $f(R,\phi)$, our model adopts the following strategy: during inflation, $f$ grows as a power law with the scale factor; during reheating, $f$ gradually decays back to its pre-inflationary value ($f\to 1$), confining the strong-coupling regime to a controllable range. According to our setup, $\alpha>0$ and $\beta<0$, which ensures that $f$ increases during inflation and decreases during reheating. To simplify the notation, we define the dimensionless ratios
\begin{equation}
\label{e34}
\epsilon_{f} \equiv \left(\frac{12 H_{\rm f}^2}{M_{\rm Pl}^2}\right)^n,
\qquad
\epsilon_{r} \equiv \left(\frac{3 H_{\rm f}^2}{M_{\rm Pl}^2}\right)^n
\end{equation}
which are both much smaller than unity in the early universe.

Setting the mass parameter $M$ to the Planck mass $M_{\rm Pl}$ and using the definitions Eqs.~\eqref{e2} and \eqref{e32}--\eqref{e34}, the coupling function can be expressed as
\begin{equation}
\label{e35}
f(R,\phi) \simeq
\begin{cases}
\left(\dfrac{a}{a_{\rm i}}\right)^{\alpha}
\left(1 - \epsilon_{f}\right)^{1/2}
& a_{\rm i}\le a \le a_{\rm f}\\
\left(\dfrac{a_{\rm f}}{a_{\rm i}}\right)^{\alpha}
\left(1 - \epsilon_{r}
\left(\dfrac{a}{a_{\rm f}}\right)^{-3n}\right)^{1/2}
\left(\dfrac{a}{a_{\rm f}}\right)^{\beta}
& a_{\rm f}\le a \le a_{\rm r}\\
1
& a \ge a_{\rm r}
\end{cases}
\end{equation}
where $\alpha \equiv \nu_1/2$ and $\beta \equiv \nu_2/2$.

\subsection{Dynamics during Reheating}

Introducing the dimensionless amplitude $\mathcal{A}(\eta, k) \equiv a(\eta) f(R,\phi) A(\eta, k)$, the mode equation is Eq.~\eqref{e6}.During the inflationary stage, the asymptotic solution beyond the horizon limit is given by an expansion of Bessel functions, which can be written as
\begin{equation}
\mathcal{A}_{\text{f}}(k,\eta) = \sqrt{-k\eta}\Bigl[ d_1(k) J_{-\alpha-\frac12}(-k\eta) + d_2(k) J_{\alpha+\frac12}(-k\eta) \Bigr]
\label{e36}
\end{equation}
with
\begin{equation}
d_1=\sqrt{\frac{\pi}{4k}}\,\frac{e^{\frac{i\pi\alpha}{2}}}{\cos(-\pi\alpha)},\quad
d_2=\sqrt{\frac{\pi}{4k}}\,\frac{e^{\frac{i\pi(-\alpha+1)}{2}}}{\cos(-\pi\alpha)}
\label{e37}
\end{equation}
During the reheating stage, the scale factor and the coupling function are given by Eq.~\eqref{e32} and Eq.~\eqref{e35}, respectively. The general solution in the long-wavelength limit $k \ll aH$ is
\begin{equation}
\label{e38}
\bar{A}_{\text{r}}(k \ll aH, \eta) = D_1(k) + D_2(k) \int_{\eta_{\text{f}}}^{\eta} \frac{1}{f_{\text{r}}^2(\bar{\eta})} d\bar{\eta}
\end{equation}
where $D_1(k)$ and $D_2(k)$ are determined by the Bunch--Davies initial conditions. $D_1(k)$ corresponds to the decaying mode, while $D_2(k)$ corresponds to the growing mode.

To determine the integration constants $D_1(k)$ and $D_2(k)$ in the mode function during the reheating stage, we impose matching conditions at the end of inflation $\eta = \eta_{\rm f}$. Defining the renormalized amplitude $\bar{A}(\eta,k) \equiv \mathcal{A}(\eta,k)/f(R,\phi)$, which satisfies its respective equation of motion in both the inflationary and reheating epochs, the coefficients are obtained as~\cite{2024_papanikolaou}
\begin{equation}
\label{e39}
\begin{aligned}
D_1(k) &\simeq
\left[1-\epsilon_f\right]^{-1/2}
\left(\frac{k}{H_{\text{f}}}\right)^{-\alpha}
\Bigg[
\frac{2^{\frac12+\alpha}d_1}{\Gamma\left(\frac12-\alpha\right)} -\frac{2^{-\frac32+\alpha}d_1 k^2}{a_{\text{f}}^2 H_{\text{f}}^2\,\Gamma\left(\frac32-\alpha\right)}
+\frac{2^{-\frac92+\alpha}d_1 k^4}{a_{\text{f}}^4 H_{\text{f}}^4\,\Gamma\left(\frac52-\alpha\right)} \\
&\quad +\frac{2^{-\frac12-\alpha}d_2\left(\frac{k}{a_{\text{f}}H_{\text{f}}}\right)^{1+2\alpha}}{\Gamma\left(\frac32+\alpha\right)}
\Bigg] \\
D_2(k) &\simeq\left[1-\epsilon_f\right]^{1/2}
\Bigg[2^{-\frac52-\alpha}\,a_{\text{f}}^{2\alpha-3}
\left(\frac{k}{H_{\text{f}}}\right)^{1-\alpha}\Bigg]\Bigg[-\frac{8\alpha+4}{\Gamma\left(\frac32+\alpha\right)}
\,a_{\text{f}}^3 H_{\text{f}} d_2
\left(\frac{k}{a_{\text{f}}H_{\text{f}}}\right)^{2\alpha} \\
&\quad -\frac{2^{2\alpha}}{\Gamma\left(\frac52-\alpha\right)}d_1 k^3 H_{\text{f}}^{-2}
+\frac{2^{2+2\alpha}}{\Gamma\left(\frac32-\alpha\right)}d_1 a_{\text{f}}^2 k
\Bigg]
\end{aligned}
\end{equation}
After expressing $f^2(R,\phi)$ in terms of $a(\eta)$, the integral can be performed analytically. Using $d\eta = da/(a^2H)$ together with the expression for the reheating epoch in Eq.~\eqref{e35}, and keeping only the dominant contributions, the integral in Eq.~\eqref{e38} reduces to a combination of power-law functions of $a/a_{\rm f}$. The fields evolve according to powers $(-2\beta+1/2)$ and $(-2\beta-3n+1/2)$. Finally, through case-by-case analysis, we obtain the following results:
\begin{itemize}
\item[(i)] $\beta<-1/4$ and $\beta<-\dfrac{1+6n}{4}$
\begin{equation}
\bar{A}_{\text{r}}(k \ll aH,\eta)=
D_2(k)\,\frac{1}{a_{\text{f}}^{2\alpha+1} H_{\text{f}}}
\Bigg[
\frac{\left(\frac{a}{a_{\text{f}}}\right)^{-2\beta+\frac12}-1}
{-2\beta+\frac12}+\epsilon_{r}
\frac{\left(\frac{a}{a_{\text{f}}}\right)^{-2\beta-3n+\frac12}-1}
{-2\beta-3n+\frac12}
\Bigg]
\label{e310}
\end{equation}
\item[(ii)] $- \dfrac{1+6n}{4}<\beta<-1/4$
\begin{equation}
\bar{A}_{\rm r}(k\ll aH,\eta)
= D_2(k)\,\frac{1}{a_{\text{f}}^{2\alpha+1} H_{\text{f}}}
\Bigg[
\frac{\left(\frac{a}{a_{\text{f}}}\right)^{-2\beta+\frac12}-1}
{-2\beta+\frac12}\Bigg]
\label{e311}
\end{equation}
\item[(iii)]  $\beta>-1/4$ and $\beta>- \dfrac{1+6n}{4}$($a<a_c$)
\begin{equation}
    \bar{A}_{\text{r}}(k \ll aH,\eta) = D_1(k)
    \label{e312}
\end{equation}
where $a_c$ is the critical scale factor
\begin{equation}
\label{e313}
\quad a_c = a_{f} \exp\left\{
\frac{3 a_{f}^2 H_{f}^2}
{(1-\epsilon_{r})k^2}
\right\}
\end{equation}
\end{itemize}
With $F(\eta)/f^2, G(\eta)/f^2 \to 1$, 
Eqs.~\eqref{e13} and \eqref{e14} reduce to the mode-energy spectra:
\begin{equation}
\label{e314}
\begin{aligned}
\frac{\mathrm{d}\rho_{\text{B}}(k,\eta)}{\mathrm{d}\ln k} &= \frac{1}{2\pi^2} \frac{k^5}{a^4} |\mathcal{A}(k,\eta)|^2 \\
\frac{\mathrm{d}\rho_{\text{E}}(k,\eta)}{\mathrm{d}\ln k} &= \frac{f^2}{2\pi^2} \frac{k^3}{a^4} \left| \left( \frac{\mathcal{A}}{f} \right)' \right|^2
\end{aligned}
\end{equation}
Taking the $\alpha=2$ branch, substituting Eqs.~\eqref{e39}-\eqref{e312} into Eq.~\eqref{e314}, and keeping only the dominant contributions, the post-inflationary magnetic and electric energy density spectra during the reheating stage reduce to, in different parameter regimes,
\begin{enumerate}
\item[(i)] $\beta<-1/4$ and $\beta<- \dfrac{1+6n}{4}$
\begin{equation}
\label{e315}
    \begin{aligned}
        \frac{d\rho_B(k,\eta)}{d\ln k}
&= \frac{k^4}{4\pi^2 a^4}
\left[1-\epsilon_{f}\right]
\left[1-\epsilon_{r}\left(\frac{a}{a_{\text{f}}}\right)^{-3n}\right] \left[1+\epsilon_{r}
\frac{-2\beta+\frac12}{-2\beta+\frac12-3n}
\left(\frac{a}{a_{\text{f}}}\right)^{-3n}\right]^2\\
&\quad \cdot \frac{1}{(-2\beta+\frac{1}{2})^2}\left(\frac{a}{a_{\text{f}}}\right)^{-2\beta+1} 
    \end{aligned}
\end{equation}
\begin{equation} 
\label{e316}
    \begin{aligned}
        \frac{d\rho_E(k,\eta)}{d\ln k}
&= \frac{k^2}{4\pi^2 a^4}
\left[1-\epsilon_{f}\right]
\left[1-\epsilon_{r}\left(\frac{a}{a_{\text{f}}}\right)^{-3n}\right]\, \left[1+\epsilon_{r}\left(\frac{a}{a_{\text{f}}}\right)^{-3n}\right]^2
H_{\text{f}}^2 a_{\text{f}}^2 \left(\frac{a}{a_{\text{f}}}\right)^{-2\beta}
    \end{aligned}
\end{equation}
\item[(ii)] $- \dfrac{1+6n}{4}<\beta<-1/4$
    \begin{align}
    \label{e317}
        \frac{d\rho_B(k,\eta)}{d\ln k}
= \frac{k^4}{4\pi^2 a^4}
\left[1-\epsilon_{f}\right]
\left[1-\epsilon_{r}\left(\frac{a}{a_{\text{f}}}\right)^{-3n}\right] \frac{1}{(-2\beta+\frac{1}{2})^2}\left(\frac{a}{a_{\text{f}}}\right)^{-2\beta+1} 
    \end{align}
\begin{equation}
\label{e318}
        \frac{d\rho_E(k,\eta)}{d\ln k}
= \frac{k^2}{4\pi^2 a^4}
\left[1-\epsilon_{f}\right]
\left[1-\epsilon_{r}\left(\frac{a}{a_{\text{f}}}\right)^{-3n}\right]\, H_{\text{f}}^2 a_{\text{f}}^2 \left(\frac{a}{a_{\text{f}}}\right)^{-2\beta}
\end{equation}
\item[(iii)]  $\beta>-1/4$ and $\beta>- \dfrac{1+6n}{4}$($a<a_c$)
\begin{equation}
\label{e319}
  \frac{d\rho_B(k,\eta)}{d\ln k}= \frac{9}{4\pi^2 a^4}
\left[1-\epsilon_{f}\right]^{-1}\left[1-\epsilon_{r}\left(\frac{a}{a_{\text{f}}}\right)^{-3n}\right]\, H_{\text{f}}^4 a_{\text{f}}^4 \left(\frac{a}{a_{\text{f}}}\right)^{2\beta}  
\end{equation}
\begin{equation}
\label{e320}
    \frac{d\rho_E(k,\eta)}{d\ln k}=0
\end{equation}
Under the constant-mode approximation, the combination $\mathcal{A}/f=\bar{A}\simeq D_1(k)$ becomes time-independent, leading to $\left[\mathcal{A}/f\right]' = 0$. This implies that all temporal evolution of the mode function is entirely absorbed by the coupling function $f(R,\phi)$, and the electric field loses its own dynamical source. This constitutes an unphysical scenario, incapable of describing the genuine electromagnetic fluctuations generated from quantum vacuum fluctuations during inflation. Therefore, this branch must be discarded.
\end{enumerate}

\subsection{Strong Coupling and Backreaction Constraints}
\label{sec:constraints}
To ensure the viability of the inflationary magnetogenesis model during the reheating epoch, two key physical constraints need to be satisfied: the strong coupling condition and the backreaction constraint, while the CMB isotropy requirement is treated separately.
\paragraph{Strong coupling problem}

To ensure the reliability of the perturbative treatment of the electromagnetic field, the effective charge $e_{\text{eff}} \equiv e/f$ must remain small throughout the evolution. As discussed above, this requires the coupling function $f$ to approach unity before the end of reheating, so that standard electromagnetism is restored. This imposes a constraint on the model parameters:

Let $N_f$ denote the number of e-folds from the start to the end of inflation, and $N_r$ denote the number of e-folds from the end of inflation to the reheating epoch. The ratio of scale factors is then defined as
\begin{equation}
\frac{a_{f}}{a_i} = e^{N_{f}}, \quad \frac{a_{r}}{a_{f}} = e^{N_{r}}
\label{e321}
\end{equation}
To avoid unacceptable strong coupling in our model, we impose the condition $f(a_{\text{reh}}) \approx 1$. Substituting the evolution of the coupling function Eq.~\eqref{e35} and employing the definitions in Eq.~\eqref{e321}, we obtain the following explicit parameter constraint:
\begin{equation}
4N_f + \ln\left[1 - \epsilon_{r} e^{-3nN_{r}}\right] + 2\beta N_{r} \approx 0
\label{e322}
\end{equation}
This equation provides a consistency relation that must be satisfied among the parameter $\beta$, the number of e-folds during inflation $N_f$, and the number of e-folds during reheating $N_r$. Only when these parameters fulfill this relation can the effective coupling remain weak throughout the evolution, thereby avoiding the strong coupling problem.

\paragraph{Backreaction problem}

To ensure the viability of the magnetogenesis model, the generated electromagnetic energy density must not exceed the background energy density throughout the cosmic evolution up to the end of reheating. This requirement is given by the backreaction condition:
\begin{equation}
\rho_E + \rho_B < \rho_{\phi|_r} \approx \frac{\pi^2 g_r}{30} T_{r}^4
\label{e323}
\end{equation}
where $g_r$ is the number of relativistic degrees of freedom and $T_r$ is the reheating temperature. Integrating the power spectrum over momentum modes from the infrared cutoff $k_i$ to the relevant scale $k_r$ at the end of reheating $\eta = \eta_{\text{r}}$, the backreaction constraint Eq.~\eqref{e323} can be expressed as
\begin{equation}
\label{e324}
\int_{k_{i}}^{k_{r}} \frac{d\rho_E(k)}{d\ln k} d\ln k + \int_{k_{i}}^{k_{r}} \frac{d\rho_B(k)}{d\ln k} d\ln k < \frac{\pi^2 g_r}{30} T_{r}^4
\end{equation}
The modes exiting the horizon during inflation span the range from $k_i = a_i H_f$ to $k_f = a_f H_f$. Following inflation, a subset of these modes re-enters the horizon during the matter-dominated epoch. Specifically, the mode re-entering precisely at the onset of reheating is denoted by $k_r = a_r H_r$. 

Using the definition of the number of e-folds Eq.~\eqref{e321}, substituting the derived energy spectrum expressions  Eqs.\eqref{e315}--\eqref{e318} into the inequality Eq.~\eqref{e324}, and taking the logarithm, we obtain the following specific constraints on the model parameters:
\begin{enumerate}
\item[(i)] $\beta<-1/4$ and $\beta<- \dfrac{1+6n}{4}$
\begin{equation}
(5 + 2\beta)N_{r} > 4\ln\frac{H_{f}}{T_{r}} - \ln\frac{\pi^2 g_r}{30(Q_1 + Q_2)}
\label{e325}
\end{equation}
where,
 \begin{align}
     Q_1 &= \frac{1}{8\pi^2}
\left[1-\epsilon_f\right]
\left[1-\epsilon_r e^{-3n N_r}\right]\left[1+\epsilon_r e^{-3n N_r}\right]
\label{e53} \\ 
Q_2 &= \frac{1}{16\pi^2}\frac{1}{\left(-2\beta+\frac12\right)^2}
\left[1-\epsilon_f\right]
\left[1-\epsilon_r e^{-3n N_r}\right] \left[1+\epsilon_r
\frac{-2\beta+\frac12}{-2\beta+\frac12-3n}
e^{-3n N_r}\right]^2
\label{e326}
\end{align}  
\item[(ii)] $- \dfrac{1+6n}{4}<\beta<-1/4$
\begin{equation}
\label{e327}
\ln\left(N_f-\frac12 N_r\right)-9N_r< \ln\frac{\pi^2 g_r}{30}-4\ln\frac{H_f}{T_r}
-\ln\left(Q_1 Q_3\right)
\end{equation}
where,
 \begin{equation}
 \label{e328} 
     Q_3 = \frac{9}{4\pi^2}\left[1-\epsilon_f\right]^{-1}
\left[1-\epsilon_r e^{-3n N_r}\right]     
 \end{equation}
\end{enumerate}
This inequality imposes strict restrictions on the parameter $\beta$ and the reheating e-folds number $N_r$. For a successful magnetogenesis scenario, the parameters must simultaneously satisfy this backreaction constraint.

\paragraph{CMB isotropy problem}

To account for the isotropy of the cosmic microwave background (CMB), the comoving horizon at the onset of inflation must be sufficiently large to encompass the scale of the present-day observable universe. This physical requirement translates into a constraint on the scale factor and the inflationary energy scale:
\begin{equation}
(a_0 H_0)^{-1} < (a_i H_i)^{-1} \quad \Longrightarrow \quad \frac{1}{H_0}\frac{a_r a_f a_i}{a_0 a_r a_f} < \frac{1}{H_f}
\label{e57}
\end{equation}
Using Eq.~\eqref{e321} together with the entropy conservation relation during the radiation-dominated era $a_0/a_r = (g_r/g_0)^{1/3} T_r/T_0$, the above condition can be rewritten as a lower bound on the total number of e-folds:
\begin{equation}
N_f + N_r > \ln\left(\frac{T_0}{H_0}\right) - \ln\left(\frac{T_r}{H_f}\right) - \frac{1}{3}\ln\left(\frac{g_r}{g_0}\right)
\label{e58}
\end{equation}
Furthermore, the number of e-folds during the reheating stage, $N_r$, can be determined by the ratio of the energy density at the end of inflation, $\rho_{\text{inf}}$, to the energy density at the end of reheating, $\rho_{\phi}|_r$. Assuming that the universe subsequently enters the standard radiation-dominated era, we obtain:
\begin{equation}
N_r = \frac{1}{3} \ln \left( \frac{\rho_{\text{inf}}}{\rho_{\phi}|_r} \right) = \frac{1}{3} \ln \left[ \frac{90 H_f^2}{8\pi G \pi^2 g_r T_r^4} \right]
\label{e59}
\end{equation}
Substituting the expression for $N_r$ into the horizon constraint further restricts the model parameter space consistent with CMB observations.

Furthermore, the allowed range of the parameter $\beta$ is restricted not only by the constraints derived above but also by the choice of $n$. For instance, for $n=2$ we require $\beta < -3.25$ or $-3.25 < \beta < -0.25$, while for $n=3$ we require $\beta < -4.75$ or $-4.75 < \beta < -0.25$.

Once the model parameter space satisfies the stringent inequality constraints Eqs.~\eqref{e321}--\eqref{e58} arising from backreaction and strong coupling suppression, the inflationary magnetogenesis mechanism can maintain the validity of perturbation theory in the post-inflationary era. Specifically, for a given reheating temperature $T_r$ and inflationary energy scale $H_f$, these boundary conditions confine the parameters $\beta$ and $N_f$ within a specific allowed range. Within this self-consistent parameter window, we can further solve Eq.~\eqref{e59} for the number of e-folds during reheating $N_r$, and then, combined with the horizon exit condition, compute the primordial magnetic field coherence length $L_{c0}$ and field strength $B_0$ at the reheating epoch, thereby providing reliable initial conditions for subsequent nonlinear MHD evolution.

\subsection{Numerical Results and Physical Predictions}
\begin{table*}[ht]
\footnotesize
\centering
\caption{Predicted present-day magnetic field strength $B_0[L_{c0}]$ and coherence length $L_{c0}$ for different inflationary energy scales $H_f$ and reheating temperatures $T_r$.}
\label{tab1}
\begin{tabular*}{\textwidth}{@{\extracolsep{\fill}}ccccccc@{}}
\toprule
$H_f$ (GeV) & $T_r$ & $\beta$ & $N_f$ & $N_r$ & $L_{c0}$ (in Mpc) & $B_0[L_{c0}]$ (in G) \\
\midrule
$5.28 \times 10^{-8}$  & 5 MeV    & -3.57  & 42.94 & 24.06 & $1.51\times 10^{-4}$ & $3.56 \times 10^{-7}$ \\
$5.28 \times 10^{-8}$  & 150 MeV  & -4.26  & 40.38 & 18.95 & $3.76 \times 10^{-6}$ & $2.24 \times 10^{-7}$ \\
$5.28 \times 10^{-8}$  & 100 GeV  & -7.10  & 35.97 & 10.13 & $5.23 \times 10^{-9}$ & $1.28 \times 10^{-7}$ \\
$5.28 \times 10^{-8}$  & 1000 GeV & -9.80  & 34.42 & 7.02  & $5.14 \times 10^{-10}$ & $9.22 \times 10^{-8}$ \\
$3.67 \times 10^{-10}$ & 5 MeV    & -4.22  & 43.77 & 20.75 & $1.51 \times 10^{-4}$ & $3.05 \times 10^{-7}$ \\
$3.67 \times 10^{-10}$ & 150 MeV  & -5.27  & 41.21 & 15.63 & $3.76 \times 10^{-6}$ & $1.84 \times 10^{-7}$ \\
$3.67 \times 10^{-10}$ & 100 GeV  & -10.80  & 36.80 & 6.82  & $5.23 \times 10^{-9}$ & $8.52 \times 10^{-8}$ \\
$3.67 \times 10^{-12}$ & 1000 GeV & -19.00 & 35.25 & 3.71  & $5.14 \times 10^{-10}$ & $4.83 \times 10^{-8}$ \\
\bottomrule
\end{tabular*}
\end{table*}

To assess the viability of the proposed magnetogenesis model, we numerically estimate the present-day magnetic field strength and its coherence length. The calculations adopt standard cosmological parameters, including the current CMB temperature $T_0 = 2.725$ K, the Hubble parameter $H_0 = 67.36$ km s$^{-1}$ Mpc$^{-1}$, the reduced Planck mass $M_{\rm Pl} = 2.435 \times 10^{18}$ GeV, and the present-day number of relativistic degrees of freedom $g_0 = 3.36$.

The coherence length $L_c$ and the magnetic field strength $B[L_c]$ at the reheating epoch are respectively defined as
\begin{align}
L_c &= a_r \frac{\displaystyle\int_0^{k_r} \frac{2\pi}{k} \frac{d\rho_B(k,\eta)}{d\ln k} d\ln k}
{\displaystyle\int_0^{k_r} \frac{d\rho_B(k,\eta)}{d\ln k} d\ln k}\label{e60}\\
B[L_c] &= \sqrt{8\pi \frac{d\rho_B(k,\eta)}{d\ln k}} \; \bigg|_{k=\frac{2\pi a_r}{L_c}}\label{e61}
\end{align}

After reheating, the universe becomes a conducting relativistic plasma in which the early generated electric field rapidly dissipates. Neglecting the nonlinear magnetohydrodynamic (MHD) cascade on sub-Hubble scales for now (deferred to later discussion), the magnetic field is solely diluted by the cosmic expansion. We perform the estimation within the reheating temperature range of $T_r = 5\,\text{MeV}$ to $1000\,\text{GeV}$, covering the QCD phase transition at $150\,\text{MeV}$ and the electroweak phase transition at $100\,\text{GeV}$. During the linear cosmic expansion, the comoving coherence length and the magnetic field strength undergo redshift according to the scale factor. Evolving the physical quantities from the reheating epoch ($a_r$) to the present cosmic time ($a_0$), one obtains the present-day coherence scale $L_{c0}$ and magnetic field amplitude $B_0[L_{c0}]$:
\begin{align}
L_{c0} &= L_c \left(\frac{a_0}{a_r}\right)\label{e62} \\
B_0[L_{c0}] &= B[L_c] \left(\frac{a_0}{a_r}\right)^{-2}\label{e63}
\end{align}
Table~\ref{tab1} presents the numerical results for various combinations of the inflationary energy scale $H_f$ and the reheating temperature $T_r$ in the parameter space, including the numbers of e-folds during inflation and reheating, $N_f$ and $N_r$, as well as the present-day magnetic field strength $B_0[L_{c0}]$ and coherence length $L_{c0}$.

The results show that a lower reheating temperature $T_r$ (e.g., $5\ {\rm MeV}$) helps maintain a larger coherence length ($L_{c0}\sim 10^{-4}\ {\rm Mpc}$) with a field strength up to $\sim 10^{-7}\ {\rm G}$, whereas a higher reheating temperature (e.g., $1000\ {\rm GeV}$) leads to a significant reduction in $N_r$, concentrating the magnetic energy onto much smaller scales ($L_{c0}\sim 10^{-10}\ {\rm Mpc}$) while achieving field strengths up to the order of $10^{-8}\ {\rm G}$. This demonstrates that the model can flexibly accommodate magnetization requirements on different cosmological scales.

When incorporating the nonlinear MHD evolution of small-scale magnetic fields, the evolution follows the Alfv\'en turbulence analysis of Banerjee \& Jedamzik~\cite{2004_banerjee} (see also Ref.~\cite{2016_subramanian,2017_sharma}). The proper magnetic field $B^{\mathrm{NL}}[L_c]$ and proper coherence length $L_c^{\mathrm{NL}}$ at the radiation--matter equality epoch $a_m$ satisfy:
\begin{equation}
\label{e64}
B_0^{\text{NL}}[L_{c0}^{\text{NL}}] = B_0[L_{c0}] \left(\frac{a_m}{a_r}\right)^{-p}, \quad L_{c0}^{\text{NL}} = L_{c0} \left(\frac{a_m}{a_r}\right)^{q}
\end{equation}
In these relations, $a_m$ denotes the scale factor at radiation--matter equality, with the exponents given by $p = (n_B+3)/(n_B+5)$ and $q = 2/(n_B+5)$. The spectral index $n_B$ is determined by the scaling of the primordial magnetic energy density, $d\rho_B/d\ln k \propto k^{n_B+3}$. This power-law evolution is applied from the radiation-dominated era to the present, neglecting the logarithmic growth of $L_c^{\text{NL}}$ after equality.

For a blue magnetic spectrum with spectral index $\alpha = 2$, the numerical simulations of Brandenburg et al.~\cite{2015_brandenburg} yield the specific exponents $p = 0.5$ and $q = 0.5$, corresponding to $B_0^{\text{NL}} \propto (a_m/a_r)^{-0.5}$ and $L_{c0}^{\text{NL}} \propto (a_m/a_r)^{0.5}$. These simulation outcomes reflect the magnetic inverse transfer effect, where magnetic energy is transported from small scales to larger scales during nonlinear turbulent evolution.
\begin{table*}[ht]
\footnotesize
\centering
\caption{Present-day magnetic field strength and coherence length under the nonlinear effect (NL) and the simulation-based nonlinear evolution (S) for different inflationary energy scales $H_f$ and reheating temperatures $T_r$.}
\label{tab2}
\begin{tabular*}{\textwidth}{@{\extracolsep{\fill}}ccccccc@{}}
\toprule
$H_f$ (GeV) & $T_r$ & $\beta$ & $L_{c0}^{\rm NL}$ (in Mpc) & $B_0^{\rm NL}$ (in G) & $L_{c0}^{S}$ (in Mpc) & $B_0^{S}$ (in G) \\
\midrule
$5.28 \times 10^{-8}$  & 5 MeV    & -3.57  & $3.11\times 10^{-2}$ & $8.40\times 10^{-12}$ & 0.45 & $1.20 \times 10^{-10}$ \\
$5.28 \times 10^{-8}$  & 150 MeV  & -4.26  & $2.92\times 10^{-3}$ & $3.70\times 10^{-13}$ & $8.16 \times 10^{-2}$ & $1.03 \times 10^{-11}$ \\
$5.28 \times 10^{-8}$  & 100 GeV  & -7.10  & $3.73 \times 10^{-5}$ & $2.51\times 10^{-15}$ & $3.15 \times 10^{-3}$ & $2.12 \times 10^{-13}$ \\
$5.28 \times 10^{-8}$  & 1000 GeV & -9.80  & $7.97 \times 10^{-6}$ & $3.81\times 10^{-16}$  & $9.97 \times 10^{-4}$ & $4.76 \times 10^{-14}$ \\
$3.67 \times 10^{-10}$ & 5 MeV    & -4.22  & $3.11\times 10^{-2}$ & $7.20\times 10^{-12}$ & 0.45 & $1.03 \times 10^{-10}$ \\
$3.67 \times 10^{-10}$ & 150 MeV  & -5.27  & $2.92\times 10^{-3}$ & $2.98\times 10^{-13}$ & $8.16 \times 10^{-2}$ & $8.48 \times 10^{-12}$ \\
$3.67 \times 10^{-10}$ & 100 GeV  & -10.80  & $3.73 \times 10^{-5}$ & $1.67\times 10^{-13}$  & $3.15 \times 10^{-3}$ & $8.48 \times 10^{-12}$ \\
$3.67 \times 10^{-12}$ & 1000 GeV & -19.00 & $7.97 \times 10^{-6}$ & $1.99\times 10^{-16}$  & $9.97 \times 10^{-4}$ & $2.49 \times 10^{-14}$ \\
\bottomrule
\end{tabular*}
\end{table*}

Table~\ref{tab2} summarizes the numerical results for different inflationary energy scales $H_f$ and reheating temperatures $T_r$, comparing the present-day coherence length $L_{c0}^{\text{NL}}$ and magnetic field strength $B_0^{\text{NL}}$ obtained from the nonlinear MHD evolution(labeled $NL$), together with the estimates $L_{c0}^{S}$ and $B_0^{S}$ from the simulation-based evolution (labeled $S$). As expected from the Alfv\'en crossing time criterion, particularly at low reheating temperatures (e.g., $T_r = 5$ MeV), the small-scale magnetic fields enter the nonlinear regime already during the radiation-dominated era.

The analysis shows that nonlinear turbulent effects significantly alter the statistical properties of the magnetic field. The coherence length $L_{c0}^{\text{NL}}$ is substantially enhanced compared to the linear evolution expectation, reaching up to the order of $10^{-2}\,\text{Mpc}$ (for example, when $H_f = 5.28 \times 10^{-8}\,\text{GeV}$ and $T_r = 5\,\text{MeV}$, $L_{c0}^{\text{NL}}$ reaches $3.11 \times 10^{-2}\,\text{Mpc}$, far exceeding the nonlinear prediction at higher reheating temperatures). At the same time, the magnetic field strength $B_0^{\text{NL}}$ is significantly reduced compared to linear evolution due to the dispersion of energy over a larger volume, typically 1 to 2 orders of magnitude lower than the nonlinear simulation estimate $B_0^S$. For instance, in the case of QCD phase transition reheating ($T_r = 150\,\text{MeV}$) with $\alpha=2$, the coherence length after nonlinear evolution is $2.92 \times 10^{-3}\,\text{Mpc}$, and the field strength drops to $3.70 \times 10^{-13}\,\text{G}$. Both quantities show notable deviations from the predictions of the linear adiabatic decay scenario ($3.76 \times 10^{-6}\,\text{Mpc}$ and $2.24 \times 10^{-7}\,\text{G}$) and the numerical simulation results incorporating the inverse transfer effect ($8.16 \times 10^{-2}\,\text{Mpc}$ and $1.03 \times 10^{-11}\,\text{G}$).

\subsection{Model Parameter Space and Distribution of $\beta$ with Observational Constraints}

The preceding sections have discussed the specific influence of the magnetic field strength on the parameter $\beta$. To further assess the viability of this model in light of cosmological observations, we need to scan the parameter space in conjunction with specific energy scales.

\begin{figure*}[htbp]
    \centering
    \begin{subfigure}[b]{0.49\textwidth}
        \centering
        \includegraphics[width=\textwidth]{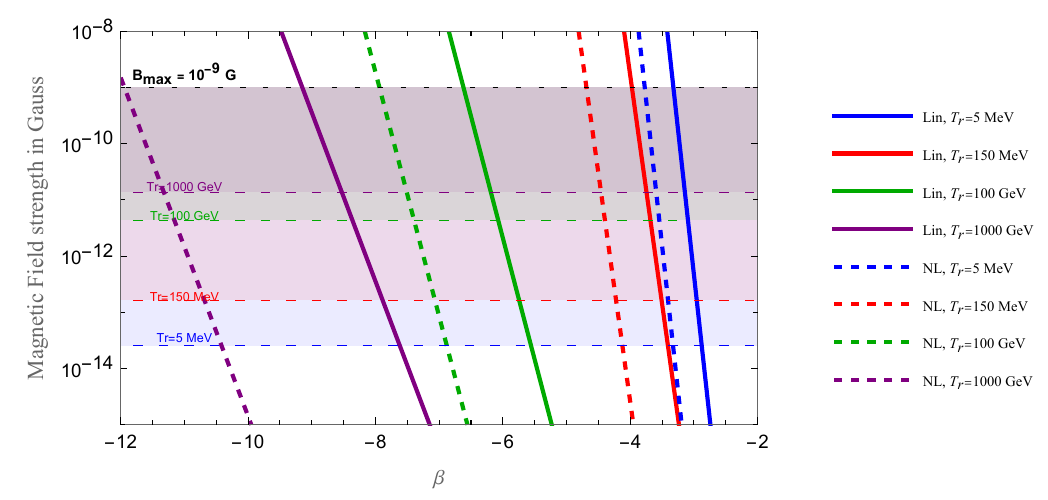}
        \caption{$H_f = 5.28 \times 10^{-8}$ GeV}
        \label{fig:sub1}
    \end{subfigure}
    \hfill
    \begin{subfigure}[b]{0.49\textwidth}
        \centering
        \includegraphics[width=\textwidth]{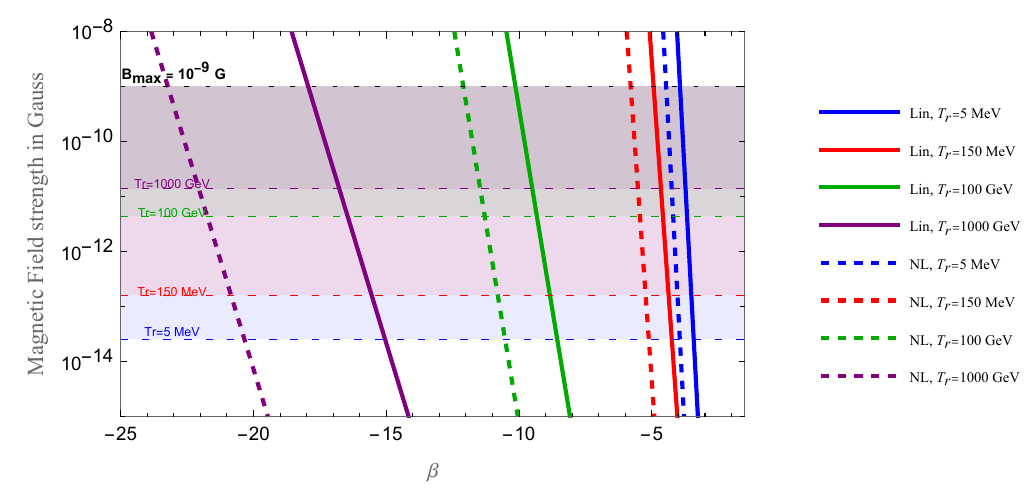}
        \caption{$H_f = 3.67 \times 10^{-10}$ GeV}
        \label{fig:sub2}
    \end{subfigure}
    \caption{Evolution of the present-day magnetic field strength $B_0$ as a function of $\beta$ for $\alpha=2$ under two inflationary energy scales. (a) $H_f = 5.28 \times 10^{-8}$ GeV; (b) $H_f = 3.67 \times 10^{-10}$ GeV. Solid lines: linear evolution; dashed lines: nonlinear evolution. Different colors correspond to different $T_r$. The colored semi-transparent bands indicate the observationally allowed regions for each $T_r$: the upper bound is set by the radio observation limit $B_{\max} = 10^{-9}\,\mathrm{G}$, and the lower bound is the reference Fermi-LAT limit $B_{\min} \approx 10^{-15}\,\mathrm{G}$ at $L_c = 0.1\,\mathrm{Mpc}$, scaled to the model's coherence length.}
    \label{fig1}
\end{figure*}
\begin{figure*}[htbp]
    \centering
    \begin{subfigure}[b]{0.49\textwidth}
        \centering
        \includegraphics[width=\textwidth]{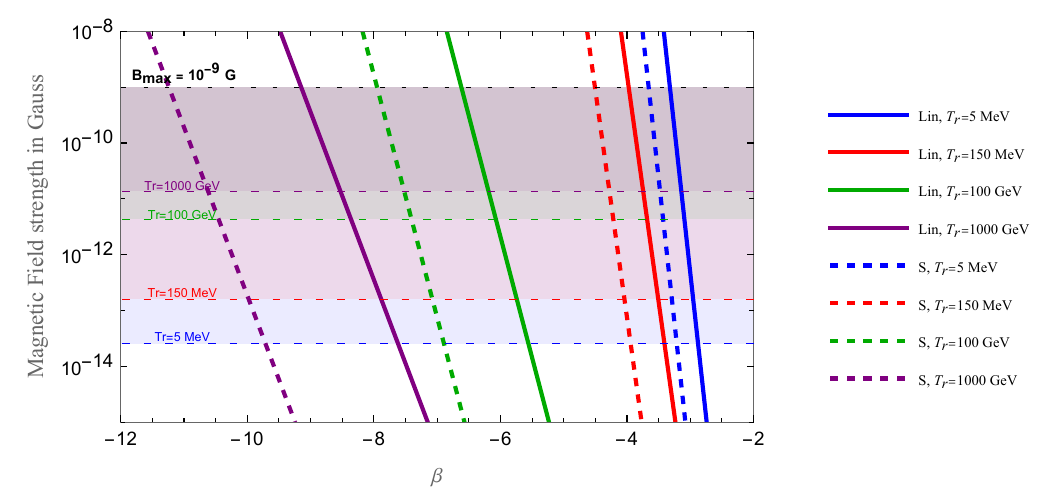}
        \caption{$H_f = 5.28 \times 10^{-8}$ GeV}
        \label{fig:sub3}
    \end{subfigure}
    \hfill
    \begin{subfigure}[b]{0.49\textwidth}
        \centering
        \includegraphics[width=\textwidth]{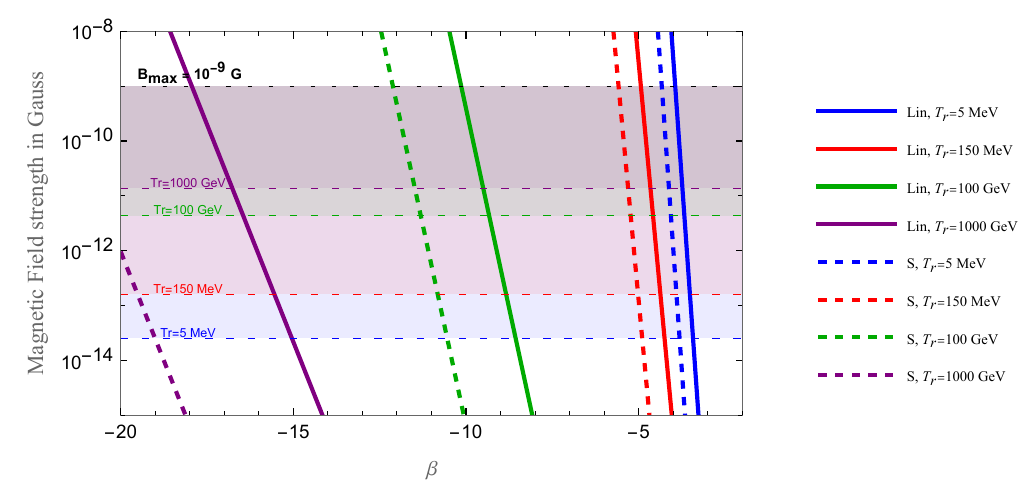}
        \caption{$H_f = 3.67 \times 10^{-10}$ GeV}
        \label{fig:sub4}
    \end{subfigure}
    \caption{Evolution of the present-day magnetic field strength $B_0$ as a function of $\beta$ for $\alpha=2$ under two inflationary energy scales. (a) $H_f = 5.28 \times 10^{-8}$ GeV; (b) $H_f = 3.67 \times 10^{-10}$ GeV. Solid lines: linear evolution; dashed lines: simulation-based nonlinear evolution. The line styles and colors follow the same definitions as in Fig.~\ref{fig1}.}
    \label{fig2}
\end{figure*}

Figure~\ref{fig1} shows the variation of the magnetic field strength with $\beta$ under two inflationary energy scales. A set of parameters $(\beta, T_r)$ is observationally allowed only when the corresponding theoretical curve (solid or dashed) passes through the colored allowed band. For example, for the low reheating temperature $T_r = 5$ MeV (blue curve), the stronger magnetic field decay arising from nonlinear turbulent evolution enables the field strength to stay above the gamma-ray lower bound and below the radio upper bound only for relatively small $\beta$ values (around $\beta \approx -3.5$). In contrast, for higher $T_r$ (e.g., the purple curve), the suppressed nonlinear effects widen the allowed parameter space.

Secondly, for a given $T_r$, nonlinear evolution (dashed lines) generally induces further attenuation of the magnetic field, leading to $B_0^{\text{NL}}$ lower than the linear estimate $B_0^{\text{Lin}}$. For a fixed parameter $\beta$, the magnetic field strength decreases significantly as $T_r$ increases. This indicates that the reheating temperature $T_r$ and the spectral index $\beta$ must satisfy stringent joint constraints, ensuring that the generated magnetic field is sufficiently strong to be potentially detectable (above the gamma-ray lower bound) while not exceeding existing observational upper limits.

In Fig.~\ref{fig2}, by comparing the linear evolution (solid lines) with the simulation-based nonlinear evolution (dashed lines), although the overall trend for the lower energy scale remains consistent with that of the higher energy scale, the curves shift noticeably to the left (toward more negative $\beta$ values). This is because a lower inflationary energy scale generates a weaker initial magnetic field, requiring a steeper spectral index (larger $|\beta|$) to compensate for cosmological redshift, thereby bringing the field strength into the observable range within the allowed bands.

We again observe that nonlinear turbulent evolution leads to significant attenuation of the magnetic field strength. Consequently, for all $T_r$, the simulation-based nonlinear evolution (dashed lines) lies systematically below the linear evolution (solid lines). For the low reheating temperature $T_r = 5$ MeV (blue curves), despite the noticeable decay from nonlinear evolution, the curve still passes through the blue observationally allowed band, indicating that this parameter combination remains viable under nonlinear evolution. However, as the reheating temperature $T_r$ increases, the overall magnetic field strength decreases, and the discrepancy between the nonlinear and linear evolution gradually grows. This indicates that for higher $T_r$, the suppression of the magnetic field strength caused by nonlinear turbulent effects becomes more prominent, imposing tighter constraints on the allowed range of $\beta$. This further confirms that accurate modeling of nonlinear evolution is crucial for assessing the viability of primordial magnetic fields.

\begin{figure*}[htbp]
    \centering
    \begin{subfigure}[b]{0.49\textwidth}
        \centering
        \includegraphics[width=\textwidth]{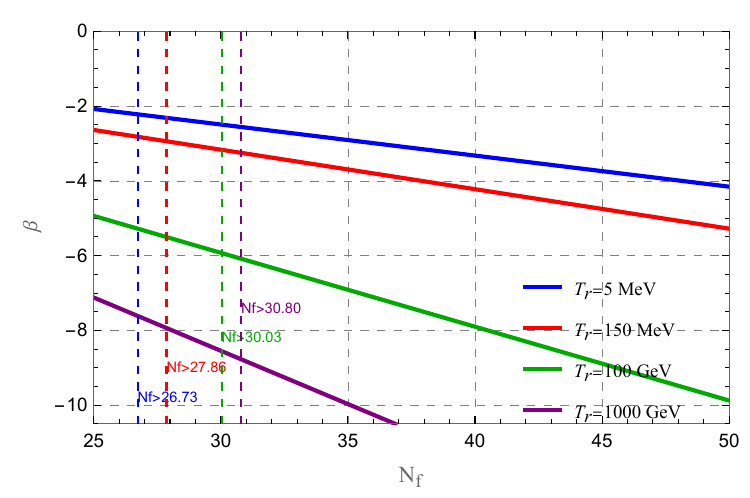}
        \caption{$H_f = 5.28 \times 10^{-8}$ GeV}
        \label{fig:sub5}
    \end{subfigure}
    \hfill
    \begin{subfigure}[b]{0.49\textwidth}
        \centering
        \includegraphics[width=\textwidth]{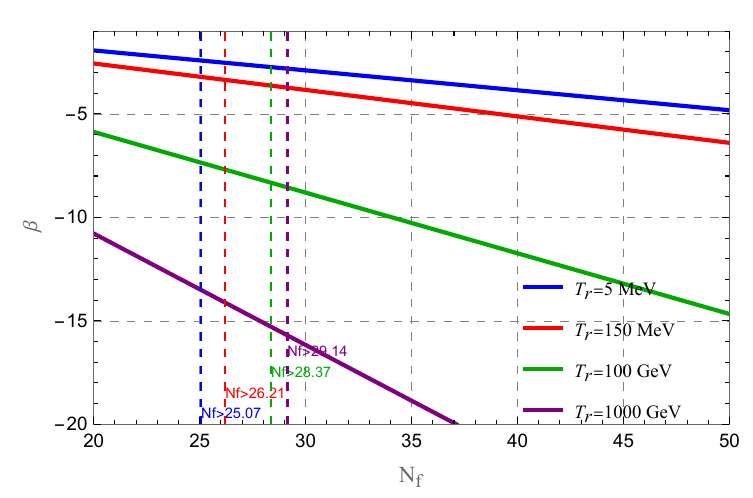}
        \caption{$H_f = 3.67 \times 10^{-10}$ GeV}
        \label{fig:sub6}
    \end{subfigure}
    \caption{ Variation of the spectral index $\beta$ as a function of the number of e-folds during inflation $N_f$, for two different inflationary energy scales: (a) $H_f = 5.28 \times 10^{-8}~\mathrm{GeV}$ and (b) $H_f = 3.67 \times 10^{-10}~\mathrm{GeV}$. The solid lines of different colors represent different reheating temperatures $T_r$ (ranging from 5 MeV to 1000 GeV). The vertical dashed lines indicate the critical positions of the required constraints (the $N_f$ threshold) for different $T_r$, with the corresponding numerical values labeled in the figure.}
    \label{fig3}
\end{figure*}

To further understand the origin of the differences in the curve shapes for different reheating temperatures $T_r$ in the $B_0$-$\beta$ relation, we show in Fig.~\ref{fig3} the evolution of the parameter $\beta$ as a function of the number of e-folds during inflation $N_f$ under two energy scales.

It can be seen that for a fixed $H_f$, the higher the reheating temperature $T_r$, the faster $\beta$ decreases with $N_f$, and the larger the critical value of the e-folds number. Moreover, when the inflationary energy scale $H_f$ is lowered, the entire $\beta$-$N_f$ curve shifts downward (i.e., $\beta$ becomes more negative), and the observationally allowed parameter space narrows accordingly.

\begin{figure*}[htbp]
    \centering
    \begin{subfigure}[b]{0.49\textwidth}
        \centering
        \includegraphics[width=\textwidth]{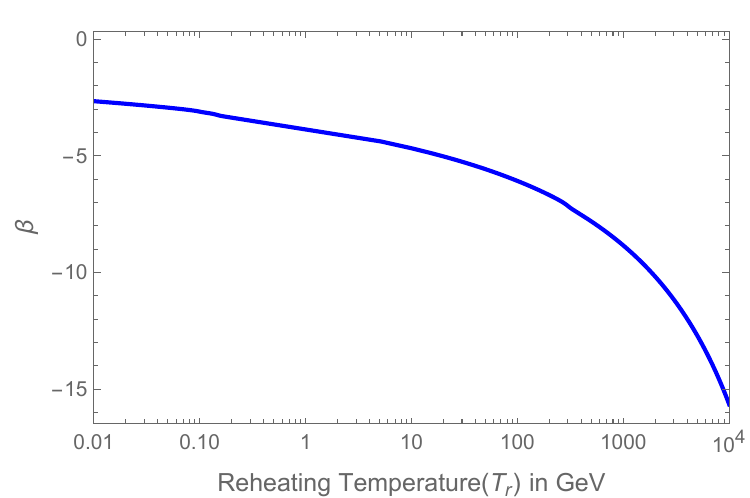}
        \caption{$H_f = 5.28 \times 10^{-8}$ GeV}
        \label{fig:sub7}
    \end{subfigure}
    \hfill
    \begin{subfigure}[b]{0.49\textwidth}
        \centering
        \includegraphics[width=\textwidth]{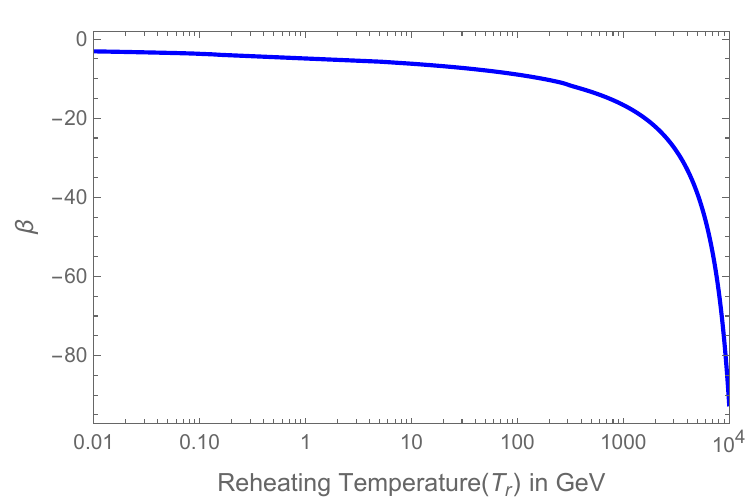}
        \caption{$H_f = 3.67 \times 10^{-10}$ GeV}
        \label{fig:sub8}
    \end{subfigure}
    \caption{ The $\beta$-$T_r$ relation for the two inflationary energy scales $H_f$. The allowed $(\beta, T_r)$ regions strictly correspond to the critical $N_f$ thresholds discussed in Fig.~\ref{fig3}, showing that higher $T_r$ and lower $H_f$ drive $\beta$ to more negative values.}
    \label{fig4}
\end{figure*}

To directly visualize how the required spectral index $\beta$ behaves under the constraint of the reheating temperature $T_r$, we present the $\beta$-$T_r$ relation in Fig.~\ref{fig4} for the two inflationary energy scales $H_f$. 

As observed in the $\beta$-$N_f$ curves of Fig.~\ref{fig3}, a lower $H_f$ shifts $\beta$ to more negative values. Consequently, in Fig.~\ref{fig4}, the curves for the lower energy scale ($H_f = 3.67 \times 10^{-10}$ GeV) are noticeably shifted toward larger $|\beta|$ compared to the higher energy scale. This reflects that a weaker initial field generated at a lower $H_f$ requires a steeper spectrum to compensate for the cosmological redshift and reach the observable range. 

In summary, combining the results from Fig.~\ref{fig1} and Fig.~\ref{fig2}, for a fixed $H_f$ and $\beta$, the magnetic field strength decreases substantially as $T_r$ increases. Comparing the linear and nonlinear evolutions, the latter introduces notable additional attenuation, shifting the viable parameter space toward more negative $\beta$. Thus, the allowed $(\beta, T_r)$ regions in Fig.~\ref{fig4} strictly correspond to the critical $N_f$ thresholds and the narrowing observationally allowed bands presented in Fig.~\ref{fig3}. These features are consistent with the behavior of the $B_0$–$\beta$ curves in Figs.~\ref{fig1} and~\ref{fig2} and the $\beta$–$N_f$ relation in Fig.~\ref{fig3}: both higher $T_r$ and lower $H_f$ drive $\beta$ toward more negative values, thereby modulating the magnitude of $B_0$ and its compatibility with observational constraints.

Within these viable parameter ranges, the primordial magnetic fields evade the strong-coupling regime while retaining sufficient amplitude to satisfy existing observational bounds. This confirms that appropriately chosen combinations of $T_r$ and $\beta$ can generate observable primordial magnetic fields without encountering theoretical inconsistencies.

\section{Summary and Discussion\label{s4}}

In this paper, we have systematically investigated the inflationary magnetogenesis mechanism within the framework of curvature-dependent coupling $f(R,\phi)$, with particular emphasis on the post-inflationary dynamics during the reheating epoch and the associated observational constraints.

Since the standard electromagnetic action is conformally invariant, conformal invariance must be broken for inflationary magnetogenesis to operate. We achieve this by introducing a coupling of the form $K(R)\, I^2(\phi)\, F_{\mu\nu}F^{\mu\nu}$ in the Lagrangian, where $K(R) = 1 - (R/M^2)^n$. This construction extends the Ratra model to a general spacetime background. We derive the modified Maxwell equations and obtain analytic expressions for the magnetic and electric energy density spectra via the near-de Sitter approximation.

Two branches of scale-invariant magnetic spectra are identified: $\alpha = -3$ and $\alpha = 2$. The $\alpha = -3$ branch suffers from a severe backreaction problem due to the divergent growth of the electric field energy density, while the $\alpha = 2$ branch, although avoiding backreaction, encounters the strong coupling problem. We adopt the $\alpha = 2$ branch and propose a strategy to circumvent the strong coupling problem by allowing the coupling function to decay back to unity during reheating.

The core of this work lies in the systematic analysis of the reheating dynamics. By constructing a broken-power-law evolution for the scale factor and the coupling function across the inflation-to-reheating transition, we derive the mode functions and the corresponding energy density spectra during reheating. Three key theoretical constraints are imposed on the model parameter space:

\begin{itemize}
    \item \textbf{Strong coupling condition:} The effective charge $e_{\text{eff}} = e/f$ must remain small throughout the evolution, requiring $f \to 1$ by the end of reheating. This yields a consistency relation among $\beta$, $N_f$, and $N_r$.
    \item \textbf{Backreaction constraint:} The total electromagnetic energy density must not exceed the background energy density at reheating, leading to inequalities that restrict $\beta$ and $N_r$ for given $H_f$ and $T_r$.
    \item \textbf{CMB isotropy requirement:} The comoving horizon at the onset of inflation must encompass the present-day observable universe, translating into a lower bound on the total number of e-folds $N_f + N_r$.
\end{itemize}

Through numerical calculations, we obtained theoretical predictions for the present-day magnetic field strength $B_0$ and coherence length $L_{c0}$ under various combinations of the inflationary energy scale $H_f$ and the reheating temperature $T_r$. We further incorporated the magnetohydrodynamic (MHD) turbulence effects after reheating. Compared with the linear adiabatic decay, both the analytical nonlinear evolution and the simulation-based nonlinear evolution significantly modify the magnetic field properties: the former enhances the coherence length (up to $10^{-2}\,\text{Mpc}$) while suppressing the field strength, whereas the latter captures the inverse transfer of magnetic energy to larger scales, yielding quantitatively distinct modifications. Comparing these two nonlinear treatments highlights the importance of properly modeling turbulent processes when assessing the viability of primordial magnetic fields.

Finally, by comparing the model parameter space with observational constraints from radio observations and Fermi-LAT gamma-ray data, we demonstrated that the inflationary energy scale $H_f$, the reheating temperature $T_r$, the parameter $\beta$, and the e-folding numbers $N_f$, $N_r$ must satisfy stringent joint constraints. The $\beta$-$N_f$ and $\beta$-$T_r$ relations further reveal that higher $T_r$ and lower $H_f$ both drive $\beta$ toward more negative values, consistent with the behavior observed in the $B_0$-$\beta$ curves.

In summary, we have extended the Ratra model to the $f(R,\phi)$ coupling framework and, combined with a carefully designed reheating dynamics, constructed a viable inflationary magnetogenesis model. This model can simultaneously satisfy theoretical consistency conditions and observational constraints. Moreover, within the viable parameter space, the model can evade the strong-coupling problem, thereby yielding detectable primordial magnetic fields.

\acknowledgments

This work was supported by the Fundamental Research Funds for the Central Universities of
Ministry of Education of China under Grants No. 3132018242,
the Natural Science Foundation of Liaoning Province of China under Grant No.20170520161 and the National Natural Science Foundation of China under Grant No.11447198 (Fund of theoretical physics).


\bibliographystyle{JHEP}
\bibliography{ls-ref}

@article{1994_kronberg,
	title = {Extragalactic magnetic fields},
    author  = {Kronberg, P.P.},
	DOI= {10.1088/0034-4885/57/4/001},
	journal = {Reports on Progress in Physics},
	year = {1994},
	volume = {57},
	pages = {4},
}

@misc{2024_papanikolaou,
      title={Gravitational-wave signatures of gravito-electromagnetic couplings}, 
      author={Theodoros Papanikolaou and Charalampos Tzerefos and Salvatore Capozziello and Gaetano Lambiase},
      year={2024},
      eprint={2408.17259},
      archivePrefix={arXiv},
      primaryClass={astro-ph.CO},
      url={https://arxiv.org/abs/2408.17259}, 
}

@article{2008_bamba,
	title = {Inflation and late-time cosmic acceleration in non-minimal {Maxwell}- \textit{{F}} ( \textit{{R}} ) gravity and the generation of large-scale magnetic fields},
	volume = {2008},
	issn = {1475-7516},
	doi = {10.1088/1475-7516/2008/04/024},
	language = {en},
	number = {04},
	journal = {Journal of Cosmology and Astroparticle Physics},
	author = {Bamba, Kazuharu and Odintsov, Sergei D},
	year = {2008},
	pages = {024},
}

@ARTICLE{2026_barroso,
       doi = {10.1088/1475-7516/2026/01/061},
       url = {https://doi.org/10.1088/1475-7516/2026/01/061},
       year = {2026},
       month = {jan},
       publisher = {IOP Publishing},
       volume = {2026},
       number = {01},
       pages = {061},
       author = {Barroso Varela, Miguel and Bertolami, Orfeu and Mantziris, Andreas},
       title = {Inflationary dynamics of non-minimally coupled f(R) matter-curvature theories},
       journal = {Journal of Cosmology and Astroparticle Physics}
}

@article{2017_sharma,
  title = {Challenges in inflationary magnetogenesis: Constraints from strong coupling, backreaction, and the Schwinger effect},
  author = {Sharma, Ramkishor and Jagannathan, Sandhya and Seshadri, T. R. and Subramanian, Kandaswamy},
  journal = {Physical Review D},
  volume = {96},
  issue = {8},
  pages = {083511},
  numpages = {13},
  year = {2017},
  month = {Oct},
  publisher = {American Physical Society},
  doi = {10.1103/PhysRevD.96.083511},
}

@article{2009_vittoria,
       doi = {10.1088/1475-7516/2009/08/025},
year = {2009},
month = {aug},
publisher = {},
volume = {2009},
number = {08},
pages = {025},
author = {Vittoria Demozzi and Viatcheslav Mukhanov and Hector Rubinstein},
title = {Magnetic fields from inflation?},
journal = {Journal of Cosmology and Astroparticle Physics},
}

@article{2025_li,
  title = {Impact of inhomogeneous perturbations of the inflaton on the cosmological primordial magnetic field},
  author = {Li, Yu and Liu, Shuang and Wang, Hang and Wang, Yao-Chuan},
  date = {2025-07-23},
  journal = {Physical Review D},
  volume = {112},
  number = {2},
  year = {2025},
  pages = {023539},
  doi = {10.1103/qhg8-nx1s},
  url = {https://link.aps.org/doi/10.1103/qhg8-nx1s}
}

@article{2026_wang,
  title = {The equivalence between Einstein and Jordan frames: a study based on the inflationary magnetogenesis model},
  shorttitle = {The equivalence between Einstein and Jordan frames},
  author = {Wang, Hang and Liu, Shuang and Li, Yu and Wang, Yao-Chuan},
  date = {2026-01-30},
  journal = {International Journal of Modern Physics D},
  volume = {35},
  number = {02},
  year = {2026},
  pages = {2550094},
  doi = {10.1142/S0218271825500944},
  url = {https://www.worldscientific.com/doi/10.1142/S0218271825500944}
}

@article{2015_brandenburg,
  title = {Nonhelical Inverse Transfer of a Decaying Turbulent Magnetic Field},
  author = {Brandenburg, Axel and Kahniashvili, Tina and Tevzadze, Alexander G.},
  journal = {Physical Review Letters},
  volume = {114},
  issue = {7},
  pages = {075001},
  numpages = {5},
  year = {2015},
  month = {Feb},
  publisher = {American Physical Society},
  doi = {10.1103/PhysRevLett.114.075001},
}

@article{2013_giovannini,
   title = {Inflationary susceptibilities, duality, and large-scale magnetic field generation},
   volume = {88},
   ISSN = {1550-2368},
   year = {2013},
   doi = {10.1103/physrevd.88.083533},
   number = {8},
   journal = {Physical Review D},
   publisher = {American Physical Society (APS)},
   author = {Giovannini, Massimo}
    }

@article{1999_blasi,
	title = {Cosmological Magnetic Field Limits in an Inhomogeneous Universe},
	author = {Blasi, Pasquale and Burles, Scott and Olinto, Angela V.},
	year = {1999},
	journal = {The Astrophysical Journal},
	volume = {514},
	number = {2},
	pages = {L79},
	doi = {10.1086/311958}
}

@article{2010_neronov,
	title = {Evidence for Strong Extragalactic Magnetic Fields from Fermi Observations of TeV Blazars},
	author = {Neronov, Andrii and Vovk, Ievgen},
	year = {2010},
	journal = {Science},
	volume = {328},
	number = {5974},
	pages = {73--75},
	doi = {10.1126/science.1184192}
}

@article{2020_osullivan,
	title = {New constraints on the magnetization of the cosmic web using LOFAR Faraday rotation observations},
	author = {O'Sullivan, S P and Br{\"u}ggen, M and Vazza, F and Carretti, E and Locatelli, N T and Stuardi, C and Vacca, V and Vernstrom, T and Heald, G and Horellou, C and Shimwell, T W and Hardcastle, M J and Tasse, C and R{\"o}ttgering, H},
	year = {2020},
	journal = {Monthly Notices of the Royal Astronomical Society},
	volume = {495},
	number = {3},
	pages = {2607--2619},
	doi = {10.1093/mnras/staa1395}
}

@article{2020_tiede,
	title = {Constraints on the Intergalactic Magnetic Field from Bow Ties in the Gamma-Ray Sky},
	author = {Tiede, Paul and Broderick, Avery E. and Shalaby, Mohamad and Pfrommer, Christoph and Puchwein, Ewald and Chang, Philip and Lamberts, Astrid},
	year = {2020},
	journal = {The Astrophysical Journal},
	volume = {892},
	number = {2},
	pages = {123},
	doi = {10.3847/1538-4357/ab737e}
}

@article{2002_widrow,
	title = {Origin of galactic and extragalactic magnetic fields},
	author = {Widrow, Lawrence M.},
	year = {2002},
	journal = {Reviews of Modern Physics},
	volume = {74},
	number = {3},
	pages = {775--823},
	doi = {10.1103/RevModPhys.74.775}
}

@article{2005_hanayama,
	title = {Biermann Mechanism in Primordial Supernova Remnant and Seed Magnetic Fields},
	author = {Hanayama, Hidekazu and Takahashi, Keitaro and Kotake, Kei and Oguri, Masamune and Ichiki, Kiyotomo and Ohno, Hiroshi},
	year = {2005},
	journal = {The Astrophysical Journal},
	volume = {633},
	number = {2},
	pages = {941},
	doi = {10.1086/491575}
}

@article{2018_safarzadeh,
	title = {Primordial black holes as seeds of magnetic fields in the universe},
	author = {Safarzadeh, Mohammadtaher},
	year = {2018},
	journal = {Monthly Notices of the Royal Astronomical Society},
	volume = {479},
	number = {1},
	pages = {315--318},
	doi = {10.1093/mnras/sty1486}
}

@article{2021_araya,
	title = {Magnetic field generation from PBH distributions},
	author = {Araya, I J and Rubio, M E and San~Mart{\'i}n, M and Stasyszyn, F A and Padilla, N D and Maga{\~n}a, J and Sureda, J},
	year = {2021},
	journal = {Monthly Notices of the Royal Astronomical Society},
	volume = {503},
	number = {3},
	pages = {4387--4399},
	doi = {10.1093/mnras/stab729}
}

@article{2023_papanikolaou,
	title = {Constraining supermassive primordial black holes with magnetically induced gravitational waves},
	author = {Papanikolaou, Theodoros and Gourgouliatos, Konstantinos N.},
	year = {2023},
	journal = {Physical Review D},
	volume = {108},
	number = {6},
	pages = {063532},
	doi = {10.1103/PhysRevD.108.063532}
}

@article{2023_papanikolaoua,
	title = {Primordial magnetic field generation via primordial black hole disks},
	author = {Papanikolaou, Theodoros and Gourgouliatos, Konstantinos N.},
	year = {2023},
	journal = {Physical Review D},
	volume = {107},
	number = {10},
	pages = {103532},
	doi = {10.1103/PhysRevD.107.103532}
}

@article{2017_archambault,
	title = {Search for Magnetically Broadened Cascade Emission from Blazars with VERITAS},
	author = {Archambault, S. and others},
	year = {2017},
	journal = {The Astrophysical Journal},
	volume = {835},
	number = {2},
	pages = {288},
	doi = {10.3847/1538-4357/835/2/288}
}

@article{2018_ackermann,
	title = {The Search for Spatial Extension in High-latitude Sources Detected by the Fermi Large Area Telescope},
	author = {Ackermann, M. and others},
	year = {2018},
	journal = {The Astrophysical Journal Supplement Series},
	volume = {237},
	number = {2},
	pages = {32},
	doi = {10.3847/1538-4365/aacdf7}
}

@article{2004_banerjee,
	title = {Evolution of cosmic magnetic fields: From the very early Universe, to recombination, to the present},
	shorttitle = {Evolution of cosmic magnetic fields},
	author = {Banerjee, Robi and Jedamzik, Karsten},
	year = {2004},
	journal = {Physical Review D},
	volume = {70},
	number = {12},
	pages = {123003},
	doi = {10.1103/PhysRevD.70.123003}
}

@article{1988_turner,
	title = {Inflation-produced, large-scale magnetic fields},
	author = {Turner, Michael S. and Widrow, Lawrence M.},
	year = {1988},
	journal = {Physical Review D},
	volume = {37},
	number = {10},
	pages = {2743--2754},
	doi = {10.1103/PhysRevD.37.2743}
}

@article{1992_ratra,
	title = {Cosmological ``Seed'' Magnetic Field from Inflation},
	author = {Ratra, Bharat},
	year = {1992},
	journal = {The Astrophysical Journal},
	volume = {391},
	pages = {L1},
	doi = {10.1086/186384}
}

@article{2008_martin,
	title = {Generation of large scale magnetic fields in single-field inflation},
	author = {Martin, J{\'e}r{\^o}me and Yokoyama, Jun'ichi},
	year = {2008},
	journal = {Journal of Cosmology and Astroparticle Physics},
	volume = {2008},
	number = {01},
	pages = {025},
	doi = {10.1088/1475-7516/2008/01/025}
}

@article{2014_kobayashi,
	title = {Schwinger effect in 4D de Sitter space and constraints on magnetogenesis in the early universe},
	author = {Kobayashi, Takeshi and Afshordi, Niayesh},
	year = {2014},
	journal = {Journal of High Energy Physics},
	volume = {2014},
	number = {10},
	pages = {166},
	doi = {10.1007/JHEP10(2014)166}
}

@article{2015_giovannini,
	title = {Inflationary magnetogenesis, derivative couplings, and relativistic Van der Waals interactions},
	author = {Giovannini, Massimo},
	year = {2015},
	journal = {Physical Review D},
	volume = {92},
	number = {4},
	pages = {043521},
	doi = {10.1103/physrevd.92.043521}
}

@article{2015_tasinato,
	title = {A scenario for inflationary magnetogenesis without strong coupling problem},
	author = {Tasinato, Gianmassimo},
	year = {2015},
	journal = {Journal of Cosmology and Astroparticle Physics},
	volume = {2015},
	number = {03},
	pages = {040},
	doi = {10.1088/1475-7516/2015/03/040}
}

@article{2016_domenech,
	title = {Inflationary magnetogenesis with broken local U(1) symmetry},
	author = {Dom{\`e}nech, Guillem and Lin, Chunshan and Sasaki, Misao},
	year = {2016},
	journal = {Europhysics Letters},
	volume = {115},
	number = {1},
	pages = {19001},
	doi = {10.1209/0295-5075/115/19001}
}

@article{2018_sharma,
	title = {Generation of helical magnetic field in a viable scenario of inflationary magnetogenesis},
	author = {Sharma, Ramkishor and Subramanian, Kandaswamy and Seshadri, T. R.},
	year = {2018},
	journal = {Physical Review D},
	volume = {97},
	number = {8},
	pages = {083503},
	doi = {10.1103/PhysRevD.97.083503}
}

@article{2019_shtanov,
	title = {Inflationary Magnetogenesis with Helical Coupling},
	author = {Shtanov,  Yu. V. and Pavliuk, M. V.},
	year = {2019},
	journal = {Ukrainian Journal of Physics},
	volume = {64},
	number = {11},
	pages = {1009--1009},
	doi = {10.15407/ujpe64.11.1009}
}

@article{2021_bamba,
	title = {Inflationary magnetogenesis with reheating phase from higher curvature coupling},
	author = {Bamba, Kazuharu and Elizalde, E. and Odintsov, S.D. and Paul, Tanmoy},
	year = {2021},
	journal = {Journal of Cosmology and Astroparticle Physics},
	volume = {2021},
	number = {04},
	pages = {009},
	doi = {10.1088/1475-7516/2021/04/009}
}

@article{2021_giovannini,
	title = {Inflationary magnetogenesis in the perturbative regime},
	author = {Giovannini, Massimo},
	year = {2021},
	journal = {Classical and Quantum Gravity},
	volume = {38},
	number = {13},
	pages = {135018},
	doi = {10.1088/1361-6382/abf899}
}

@article{2022_durrer,
	title = {Magnetogenesis in Higgs-Starobinsky inflation},
	author = {Durrer, R. and Sobol, O. and Vilchinskii, S.},
	year = {2022},
	journal = {Physical Review D},
	volume = {106},
	number = {12},
	pages = {123520},
	doi = {10.1103/PhysRevD.106.123520}
}

@misc{2023_velasquez,
	title = {About Jordan and Einstein frames: a study in inflationary magnetogenesis},
	shorttitle = {About Jordan and Einstein frames},
	author = {Vel{\'a}squez, Joel and Hortua, H{\'e}ctor J. and Casta{\~n}eda, Leonardo},
	year = {2023},
	number = {arXiv:2303.01301},
	eprint = {2303.01301},
	primaryclass = {gr-qc},
	archiveprefix = {arXiv}
}

@misc{2025_dimopoulos,
      title={Is inflationary magnetogenesis sensitive to the post-inflationary history ?}, 
      author={Konstantinos Dimopoulos and Anish Ghoshal and Theodoros Papanikolaou},
      year={2025},
      eprint={2412.10367},
      archivePrefix={arXiv},
      primaryClass={astro-ph.CO},
      url={https://arxiv.org/abs/2412.10367}, 
}

@article{2022_li,
	title = {Inflationary magnetogenesis with a self-consistent coupling function},
	author = {Li, Yu and Zhang, Le-Yao},
	year = {2022},
	journal = {Modern Physics Letters A},
	volume = {37},
	number = {10},
	pages = {2250069},
	doi = {10.1142/S0217732322500699}
}

@article{2023_li,
	title = {Inflationary magnetogenesis of primordial magnetic fields with multiple vector fields},
	author = {Li, Yu and Zhang, Le-Yao},
	year = {2023},
	journal = {Modern Physics Letters A},
	volume = {38},
	number = {10n11},
	pages = {2350062},
	doi = {10.1142/S0217732323500621}
}

@article{2010_subramanian,
	title = {Magnetic fields in the early Universe},
	author = {Subramanian, K.},
	year = {2010},
	journal = {Astronomische Nachrichten},
	volume = {331},
	number = {1},
	pages = {110--120},
	doi = {10.1002/asna.200911312}
}

@article{2011_kandus,
	title = {Primordial magnetogenesis},
	author = {Kandus, Alejandra and Kunze, Kerstin E. and Tsagas, Christos G.},
	year = {2011},
	journal = {Physics Reports},
	volume = {505},
	number = {1},
	pages = {1--58},
	doi = {10.1016/j.physrep.2011.03.001}
}

@article{2016_subramanian,
	title = {The origin, evolution and signatures of primordial magnetic fields},
	author = {Subramanian, Kandaswamy},
	year = {2016},
	journal = {Reports on Progress in Physics},
	volume = {79},
	number = {7},
	pages = {076901},
	doi = {10.1088/0034-4885/79/7/076901}
}

%

%
%
%
%
%
%
%
%
\end{document}